\documentclass[pra,amsmath,amssymb,twocolumn,superscriptaddress]{revtex4}
\usepackage{graphicx}
\usepackage{color}
\newcommand{\halfsq}{\frac{1}{\sqrt{2}}}
\newcommand{\half}{\frac{1}{2}}
\newcommand{\qrtr}{\frac{1}{4}}

\begin{document}

\title{Counterfactual quantum certificate authorization} 
\author{Akshata Shenoy H.}
\email{akshata@ece.iisc.ernet.in}
\affiliation{Department of Electrical Communication Engineering, Indian Institute of Science, Bangalore, India}
\author{R. Srikanth}
\email{srik@poornaprajna.org}
\affiliation{PPISR, Bangalore, India}
\affiliation{Raman Research Institute, Bangalore, India}
\author{T. Srinivas}
\affiliation{Department of Electrical Communication Engineering, Indian Institute of Science, Bangalore, India}

\begin{abstract}
We present a multi-partite protocol in a counterfactual paradigm.
In   counterfactual  quantum   cryptography,  secure   information  is
transmitted between two spatially separated parties even when there is
no physical  travel of particles transferring  the information between
them.  We  propose here  a tripartite counterfactual  quantum protocol
for  the task  of  certificate authorization.   Here  a trusted  third
party, Alice, authenticates an entity Bob (e.g., a bank) that a client
Charlie   wishes  to   securely  transact   with.   The   protocol  is
counterfactual with  respect to either  Bob or Charlie.  We  prove its
security against a general incoherent attack, where Eve attacks single
particles. 
\end{abstract}
\maketitle
\date{}

\section{Introduction}
Suppose a client (Charlie)  wishes to undertake a business transaction
with  a bank  Bob.  Charlie  looks up  Bob's website  via  an internet
search but  is unsure of the website's  authenticity.  His transaction
requires him to securely  transmit confidential information to Bob.  A
solution  to  this frequently  encountered  problem  in e-commerce  is
\textit{certificate  authorization}  (CA)  where  Alice,  a  well-known
trusted  third party validates  Bob's website  on  request from
Charlie. Classically, this task is accomplished via digital signatures
and public-private keys \cite{rivest,loren}.

Alice,  as a  \textit{certificate authority},  has a  mutual agreement
with  a financial  firm,  whereby  the latter  provides  her with  the
current  information  about   Bob's  claimed  online  identity.   Upon
verifying  that  the  website  indeed  belongs to  Bob,  Alice  issues
certificates  in the  form  of digital  signatures and  public-private
keys, thereby validating Bob's website.  Charlie can now transact with
Bob  using the  latter's certified  public key.   Alice  keeps herself
updated  regarding  the  renewal and expiry  of certificates  and  current
information of  the certificate holders.  For example,  if Bob changes
the name of his website, the certificate issued to the website becomes
invalid.  To  resume transactions, he  needs to submit  an application
for a new certificate including legal documents supporting the change.

Here we  wish to  introduce a quantum  method to accomplish  the above
described  task   in  the   counterfactual  paradigm  which   we  call
\textit{counterfactual  quantum   certificate  authorization}  (CQCA).
Counterfactual quantum cryptography  \cite{GS99,N09,SLA+0} is based on
the  idea of  interaction-free  measurements \cite{EV93,KWM+99}, which
involves   communicating  information   even   without  the   physical
transmission of a  particle, a point that is  of foundational interest
\cite{SS2}.   Information  is  transferred  by  blocking  rather  than
transmitting a particle. While  this is also possible classically, 
in the  classical case, the  blockade results in a  particle detection
near the  blockade, whereas  in the quantum  case by virtue  of single
particle  nonlocality,  the  particle   may be detected  away  from  the
blockade,  which is the  counterfactual element  here.  Counterfactual
protocols     use    orthogonal     states    for     encoding    bits
\cite{GV95,ABD+10,KI97}.   Its security has  been analyzed  by various
authors \cite{YLC+10, ZWT12, ZWT+12, YLY+12}, and issues  related to
improving its  efficiency \cite{SW10} and  experimental realization by
others   \cite{RWW+11, XW11, XST+11, JSL11},  including   a  fully
counterfactual   version  of the Noh 2009 (N09)  protocol   using   a  Mach-Zehnder
interferometer setup  \cite{BCD+12}.  The present  authors proposed a
semicounterfactual quantum key distribution (QKD) protocol to clarify
the origin of security in the counterfactual paradigm \cite{SSS13}.

In the  proposed CQCA protocol,  Alice, in certifying Bob  to Charlie,
enables the latter two to share a secure random key.  In this respect,
the quantum  version differs from  classical CA, where Alice  plays no
role in  the secure communication  between Bob and Charlie.   Thus the
security  must  be  considered   with  respect  to  both  a  malicious
eavesdropper Eve as well Alice, who could overstep her CA role and try
to  eavesdrop on  their  transaction.  

The article  is structured as  follows: In Sec.  (\ref{sec:pro}), a
protocol for CQCA is presented.  In Sec. (\ref{sec:safe}), we prove
its security in the case of  a general incoherent attack by Eve, and a
semihonest  Alice.  In  the Sec.  (\ref{sec:discu}), we  provide a
summary and conclusions.

\section{A protocol for CA \label{sec:pro}}

\begin{figure}
\begin{center}
\includegraphics[width=8cm]{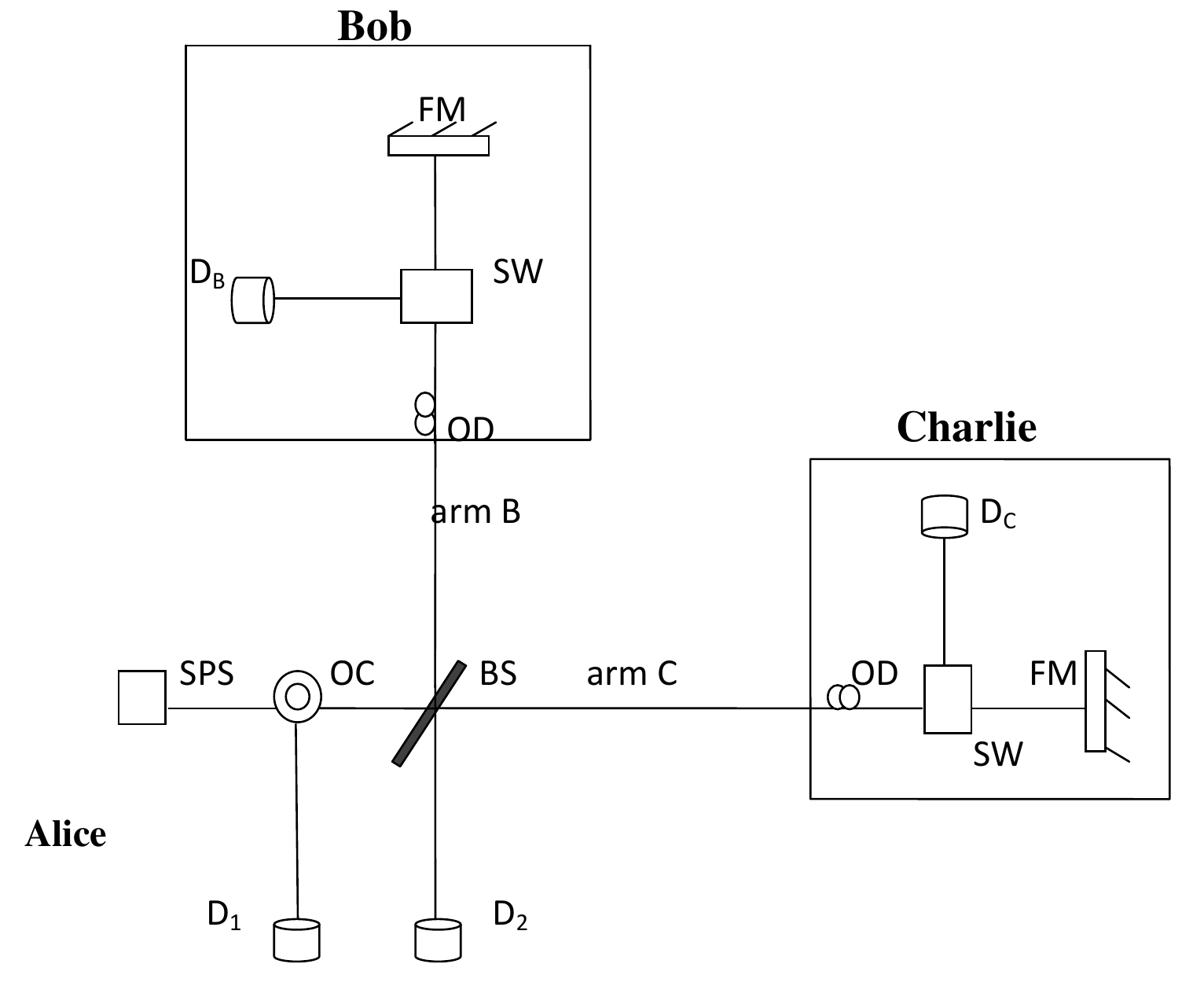}
\caption{(Color line)Experimental   set-up  for   CQCA   using  a   Michelson-type
  interferometer: Alice's module  consists of the single-photon source
  (SPS), which  initiates the protocol by sending  photons through the
  beamsplitter BS  via optical circulator OC. This  splits each photon
  into branches  along Bob's  arm ($B$) and ($C$).  The optical  delay OD
  maintains the  phase by compensating for the  path-difference in the
  two arms.   Bob (Charlie)  randomly applies either  absorption ($A$)
  using  detector $D_B$ ($D_C$)  or reflection  ($F$) using  a Faraday
  mirror.}
\label{fig:cf}
\end{center}
\end{figure}

Alice, Bob and Charlie are assumed to be online on both a conventional
classical as  well as  a quantum network.   Charlie sends  a classical
request to certificate authority Alice, whose station is equipped with
a  single-photon  source  (SPS)  and  a  beam  splitter  (BS)  (Fig.
\ref{fig:cf}).    After    classically   acknowledging   Charlie   and
classically  intimating Bob about  Charlie's contact,  Alice initiates
the protocol  on a  quantum channel by  transmitting to them  a packet
that  consists of  a single  photon,  which is  split at  BS into  the
channels that  lead to Bob (arm  $B$) and Charlie (arm  $C$). We label
these particles  $B$ and $C$.   Each transmission packet is  hybrid in
nature, consisting  of a classical  (bits) and a quantum (qubits) part,
will contain a classical  \textit{header}, a hybrid \textit{body}, and
a   possible   classical   \textit{footer}.    The   header   contains
instructions about the type of  data the packet is carrying, including
packet  length,  packet number, and the  origin  and  destination of  the
packet.  The footer consists of a  couple of bits that indicate to the
receiving device the termination of  the packet.  Thus, the header and
the footer hold control information for negotiating the network, while
the  body  will house  the  quantum data  as  well  as other  possible
conventional classical information.

A  single photon  of  arbitrary   polarization  emitted  from  SPS  is
represented after BS by
\begin{equation}
|\Psi\rangle_{BC}  = \frac{1}{\sqrt{2}}(|0\rangle_B|\psi\rangle_C  + i
|\psi\rangle_B|0\rangle_C),
\label{eq:cqca}
\end{equation}
where the first (second) ket  refers to the transmitted (reflected) or
Charlie (Bob) arm.

Bob and Charlie each  possess a photon-number resolving detector $D_B$
and $D_C$, respectively, that absorbs the photon by process $A$, and a
Faraday mirror  that applies  operation $F$, which  is to  reflect the
photon  without introducing an additional phase.   The operation  $A$ is
assumed  to  be  equipped  with  spectral  filtering  to  time-resolve
multiple photon  arrivals.  Each of the  participants randomly applies
the  operation   $F$  (reflect)   or  $A$  (absorb).    The  following
possibilities arise: (1) Bob and Charlie both apply $F$, which results
in  detection  at the  detector  $D_2$  with  probability 1.  (2)  Bob
(Charlie) applies  $F$ ($A$) or  vice versa. With  probability $\qrtr$
the particle  is detected at $D_1$  or at $D_2$,  and with probability
$\half$, it is absorbed at $D_B$ or $D_C$. (3) If Bob and Charlie both
apply $A$,  then there is necessarily  a detection at  either $D_B$ or
$D_C$.

The corresponding probabilities  are summarized in Table \ref{tab:BC}.
Bob and  Charlie adopt the convention whereby  Alice's $D_1$ detection
when they  apply $(AF)$  ($(FA)$) corresponds to  a 0 (1)  secret bit.
The  efficiency  of the  protocol  can  be  calculated as:  $P(D_1)  =
P[D_1|(F,A)] P[(F,A)] + P[D_2|(F,A)] P[(F,A)]  = (1/4) (1/4) + (1/4)
(1/4) = 1/8$.
\begin{table}[h]
\caption{Probabilities  for   outcomes  corresponding  to   Bob's  and
  Charlie's actions.}
\begin{tabular}{c|c|c}
\hline  Bob and Charlie  &  $F$  &  $A$  \\ \hline  $F$  &  ($D_2,1$)  &
($D_1,\frac{1}{4}$),           ($D_2,\frac{1}{4}$), \\
                &             &      (NULL,$\frac{1}{2}$)
\\          \hline         $A$          &         ($D_1,\frac{1}{4}$),
($D_2,\frac{1}{4}$), & (NULL,$1$) \\ 
     &   (NULL,$\frac{1}{2}$) &     \\ \hline
\end{tabular}
\label{tab:BC}
\end{table}

We  present the basic  protocol: (1)  Upon receiving  Bob's classical,
authenticated request and Charlie's  consent, Alice injects $n$ single
photons sequentially  into the input port  of the BS. (2)  Bob and Charlie
randomly  apply  operations  $F$  or  $A$  in the  arms  $B$  and  $C$,
respectively. (3) On  the $n$ outcome data collected,  a fraction $nf$
(where $f < 1$) is randomly selected by Bob and Charlie (by discussion
over an authenticated classical channel),  for which they ask Alice to
announce her detection  data (which can be NULL,  $D_1$ , or $D_2$). Bob
and Charlie announce their settings  ($A$ or $F$) and outcome (in case
of  $A$,  as to  whether  a  photon was  registered  or  not in  their
respective  detector $D_B$  or  $D_C$) information.   Bob and  Charlie
determine whether the obtained experimental data is sufficiently close
to  the  probabilities  in  Table  \ref{tab:BC}.   If  yes,  then  the
anticorrelated settings corresponding to  the $D_1$ detections form a
secure  secret  key  shared   between  them.  The  protocol  is  
counterfactual in the sense that when a secret bit is generated due to
$D_1$ detections, the photon would not have physically traveled along
one of the arms, i.e., it did not physically travel via the Bob arm or
Charlie  arm, even  though both  their choices  contribute to  the bit
generation.

(4) The closeness of the experimental data to the pattern in the table
\ref{tab:BC} is estimated using the figures of merit given below:
\begin{description}
\item[\textit{Coincidence count}]  They   verify  that   the   fraction  of
  coincidence detections when both Bob and Charlie apply $A$
\begin{equation}
\kappa \equiv P(D_BD_C|AA)
\label{eq:kappa}
\end{equation}
is sufficiently close to 0.
\item[\textit{Visibility check.}] The visibility of the interference fringes
\begin{equation}
\mathcal{V}  \equiv  \frac{P(D_2|FF)   -  P(D_1|FF)}  {P(D_1|FF)  +
  P(D_2|FF)}
\label{eq:visi0}
\end{equation}
must be sufficiently close to 1.
\item[\textit{Bias check.}]  The bias in  Alice's outcomes when  their settings
  are anti-correlated
\begin{eqnarray}
B = \textrm{max}\{\left|P(D_1|AF) - P(D_2|AF)\right|, \nonumber \\
  \left|P(D_1|FA) - P(D_2|FA)\right|\}
\label{eq:bias}
\end{eqnarray}
must be sufficiently close to 0.
\item[\textit{Determining error rate.}] The  secret bits shared between Bob and
  Charlie  are generated  precisely when  a honest  Alice  announces a
  $D_1$  detection,  for  ideally   in 
 this  case  their  inputs  are
  anti-correlated.  Deviation  from this pattern  allows them estimate
  the error rate on the raw key:
\begin{equation}
e \equiv P(FF|D_1) +  P(AA|D_1),
\label{eq:e}
\end{equation}
which must be sufficiently close to 0.
\item[\textit{Estimating multi-photon  pulses and channel  losses.}]  Two other
  figures of merit  are estimates on $r$, the  rate of multiple count,
  which  may   be  due  to   dark  counts  or   certain  photon-number
  non-preserving attacks \cite{SSS13}, and $\lambda$, transmission loss
  rate over the channel.
\end{description}

(5) In the above, if any  of $\kappa, \mathcal{V}, B, e$ and the other
figures of merit  are not sufficiently close to  their expected value,
then Bob and Charlie abort  the protocol run. Otherwise, the remaining
approximately $(1-f)n/8$ bits corresponding to Alice's $D_1$ detection
are used  for further classical  post-processing to extract  a smaller
secure  key  via key  reconciliation  and  privacy amplification. 

\section{Security \label{sec:safe}}

In classical CA, Alice only  certifies the digital signature and is by
definition  trustworthy.  By  contrast, in  the present  quantum case,
Alice participates in the key  generation.  Thus, in principle, we may
assume that she is not  to be trusted completely.  More precisely, her
action  may  be  characterized  as \textit{semi-honest}  in  that  she
fulfils her  CA role per the  official protocol, but  may collude with
Eve (Sec.  \ref{sec:AE}) to extract key information.   Our study of
the proof  of security therefore  first examines protection  against a
semihonest  Alice, while  Sec. \ref{sec:E}  considers the  case of
malicious Eve.

\subsection{Security against semihonest Alice \label{sec:AE}}

To   eavesdrop,   suppose  Alice   transmits   single  photons   along
\textit{both} arms $B$ and $C$,  and infer Bob's and Charlie's choices
deterministically according to whether the respective particle returns
to her or not.  This is foiled by the coincidence check, where Bob and
Charlie would find coincidence counts when they apply $AA$.

Alice gains  nothing by  sending photons along  one of the  arms. Even
though she gains full information on either Bob's or Charlie's choice,
she  would  know  nothing  about  the  other's  choice,  so  that  her
information   on  the   potential   secret  bit   is  nil.    Suppose,
irrationally,  that she  does  launch  such an  attack,  by sending  a
particle to Bob alone.  If she receives it back, then Bob applied $F$,
and if  not, he applied $A$.  In the latter  case, in step (3)  of the
protocol,  the only  outcome consistent  with the  experiment  is that
Alice should announce  NULL, given that Bob has  a detection. Hence no
secret bit is generated.

Now, in  the former  case, Charlie  may have applied  $F$ or  $A$ with
equal probability. Further, in  the second case, Charlie could not have
detected a particle.  If we now  consider the cases $FA$ and $FF$ such
that  Charlie did  not detect  a photon  on $D_C$,  then  Alice should
obtain    outcome   $D_1$    with    probability   $P(C{\rightarrow}A)
P(D_1|FA^\prime)     +     P(C{\rightarrow}F)P(D_1|FF)     =     \half
P(D_1|FA^\prime) +  \half P(D_1|FF) =  \half\half + 0  = \frac{1}{4}$,
and    outcome     $D_2$    with    probability    $P(C{\rightarrow}A)
P(D_2|FA^\prime)     +     P(C{\rightarrow}F)P(D_2|FF)     =     \half
P(D_2|FA^\prime)  +  \half  P(D_2|FF)  = \half\half  +  \half\cdot1  =
\frac{3}{4}$, where  $A^\prime$ denotes  that Charlie applied  $A$ and
did not detect a photon.  Now Alice needs to fake the statistics to be
compatible with the honest protocol.  Suppose Alice randomly generates
numbers 0 and 1  with probability $\frac{1}{4}$ and $\frac{3}{4}$, and
announces $D_1$  ($D_2$) when she  obtains 0 (1). Her  announcement of
$D_1$ will deterministically  lead to an error if  Charlie had applied
$F$ [since  $P(D_1|FF)=0$]. In this  fake attack, Alice does  not know
what Charlie's operation was irrespective of whether she outputs $D_1$
or $D_2$, and so $P(C\rightarrow  F|D_1) = \half$.  A similar argument
applies  if Alice sends  a particle  to Charlie  alone. Thus  if Alice
transmits  such   single-path  particles   to  Bob  or   Charlie  with
probability $p$, then  from Eq.  (5) and Table \ref{tab:BC},  we see that
Bob   and  Charlie  will   detect  an   error  with   probability  $e=
P(A\rightarrow  D_1)\half  =   \frac{p}{4}\half  =  \frac{p}{8}$.   To
counter this, Alice  may choose to announce only  $D_2$, in which case
$e=0$, but bias  $B = 2\times\frac{p}{4} = \frac{p}{2}$.  Thus such an
attack by Alice will be detected in the bias check.

\subsection{Security against Eve \label{sec:E}}

The  above  checks rule  out  Alice  from  deviating from  the  honest
protocol,  though she  may  still  collude with  Eve  (i.e., Alice  is
constrained  to be semihonest).   The last  check mentioned  above is
intended  to  guarantee that  the  SPS  and  the channel  deliver  the
required performance. Therefore, in this  analysis we do not take into
account attacks by Eve based on channel losses or imperfect sources.

We  discuss   the  security  scenario  where  Eve   attacks  each  run
individually,  by entangling  the light  along  both the  arms with  a
separate  probe positioned near  either arm.   These probes  $E_1$ and
$E_2$     are    prepared     in    the     initial     ready    state
$|R\rangle_{E_1}|R\rangle_{E_2}$.  During  the transmission from Alice
to  Bob-Charlie,  Eve   applies  the  (number-preserving)  interaction
\cite{SSS13} on the joint $BE_1$ and $CE_2$ systems:
\begin{equation}
\mathcal{K}  = |0\rangle_j\langle0|\otimes K_0  + |1\rangle_j\langle1|
\otimes K_1,
\label{eq:atakU}
\end{equation}
such that $\langle  0|K^\dag_1K_0|0\rangle \equiv \langle y|n\rangle =
\cos(\theta_j)$, where  $j \in \{B,  C\}$.  For simplicity,  we assume
$\theta_B = \theta_C = \theta$. This interaction produces the state.
\begin{eqnarray}
|\Psi^\prime\rangle_{BCE}                                           &=&
\mathcal{K}|\Psi\rangle_{BC}|RR\rangle_{E}   =  \halfsq(|\psi\rangle_B
|0\rangle_C|y,n\rangle_{E}   \nonumber   \\    &   +   &   |0\rangle_B
|\psi\rangle_C|n,y\rangle_{E}),
\label{eq:statevaK}
\end{eqnarray}
where we use  the notation $E \equiv E_1E_2$.   The Bob-Charlie action
$(F,F)$ leaves the $|\Psi\rangle_{BCE}$ unchanged.  In the case of Bob
and  Charlie  applying  $(F,A)$,  the resulting  states  are  $\halfsq
|0\rangle_B  |0\rangle_C  |n,y\rangle_{E}$  or  $\halfsq|\psi\rangle_B
|0\rangle_C |y,n\rangle_{E}$, of which the former implies detection by
Bob and the  latter leads potentially to a  $D_1$ detection for secret
bit  1.  In the  case of  $(A,F)$, the  resulting states  are $\halfsq
|0\rangle_B           |0\rangle_C           |y,n\rangle_E$          or
$\halfsq|0\rangle_B|\psi\rangle_C|n,y\rangle_E$,  of which  the former
implies detection  by Charlie  and the latter  leads potentially  to a
$D_1$  detection for  secret bit  0. The  attack does  not  affect the
probability  for secret  bit generation,  which remains,  as  in Table
\ref{tab:BC}
\begin{equation}
P(D_1|AF) = P(D_1|FA) = \frac{1}{4}
\label{eq:affad1}
\end{equation}
That Eve does not gain on attacking the return leg applies here too as
in semicounterfactual QKD \cite{SSS13}.

Thus,  the most  general  incoherent number-preserving attack  (which
entails a channel's losslessness) that Eve  can launch would be  to use
the above onward leg attack, and then measure her probe $E_1E_2$ after
Alice's announcement.  The timings of pulses transmitted by Alice must
be random, for if Eve knew  the transmission schedule, she would use an
Alice-like setup to  probe Bob's or Charlie's setting  by inserting a
photon into the stream $B$ or  $C$ in synchrony with Alice, and checks
if it  returns or not. In  principle, this trojan horse  attack can be
detected using  spectral filtering \cite{ZWT12}. An  alternative is to
exploit the  fact that coding  here is not polarizationbased,  and to
use   a  Bennett-Brassard-1984-like  \cite{bb84}   check  \cite{SSS13}.
However, the security here is undermined if Alice colludes with Eve by
supplying her with the polarization information.

In  our analysis,  we assume  the  worst-case scenario  where Eve  has
complete knowledge of the transmission schedule between Alice and Bob.
Thus she times her attack to happen just when the particle is about to
enter Bob's  station, and completes  it after Alice's  announcement of
$D_1$ detection events.

Eve jointly measures  her probes $E_1$ and $E_2$,  the information she
extracts being dependent on  her ability to distinguish between states
$|y,n\rangle_{E}$ and $|n,y\rangle_{E}$.  From Eq. (\ref{eq:statevaK}),
an upper bound on her information is the Holevo quantity
\begin{equation}
\chi(\theta)     =     S\left(\frac{\Pi_{|n,y\rangle}     +
  \Pi_{|y,n\rangle}}{2}\right)                                     -
\frac{1}{2}\left[S\left(\Pi_{|y,n\rangle}\right)               +
  S\left(\Pi_{|n,y\rangle}\right)\right],
\label{eq:holevo}
\end{equation}
where $S(\cdot)$ denotes von Neumann entropy and $\Pi_{|x\rangle}$ the
projector  to  state $|x\rangle$.   The  square-bracketed quantity  in
Eq. (\ref{eq:holevo}) vanishes because of the purity of the considered
states.     The    reduced    density    matrix   of    $E_1E_2$    in
Eq. (\ref{eq:statevaK}) is:
\begin{equation}
\rho_{E}  =  \frac{1}{2}\left( \begin{array}{cccc} 2\cos^2(\theta)  &
  \cos(\theta)\sin(\theta)          &         \cos(\theta)\sin(\theta)
  \\     \cos(\theta)\sin(\theta)     &     \sin^2(\theta)     &     0
  \\ \cos(\theta)\sin(\theta) & 0 & \sin^2(\theta)
\end{array}\right),
\end{equation}
in  the basis  $\{|y,y\rangle, |y,y^\perp\rangle,|y^\perp,y\rangle\}$,
leaving out $|y^\perp,y^\perp\rangle$,  which lies outside the support
of $\rho_{E}$.   The above  matrix is of  rank 2,  whose non-vanishing
eigenvalues  are $e_1  = \frac{1}{4}[1  - \cos(2\theta)]$  and  $e_2 =
\frac{1}{4}[3 + \cos(2\theta)]$, so that Eve's information $I_E \equiv
I_{BE} = I_{CE}$, using Eq. (\ref{eq:holevo}), is
\begin{equation}
I_E    \le    \chi(\theta)     =    H(e_1)    =    H\left(\frac{1    -
  \cos(2\theta)}{4}\right),
\label{eq:IE}
\end{equation}
where $H(x)  \equiv -x\log_2(x)-(1-x)\log_2(1-x)$ denotes  the Shannon
binary entropy.

\begin{figure}
\begin{center}
\includegraphics[width=8cm]{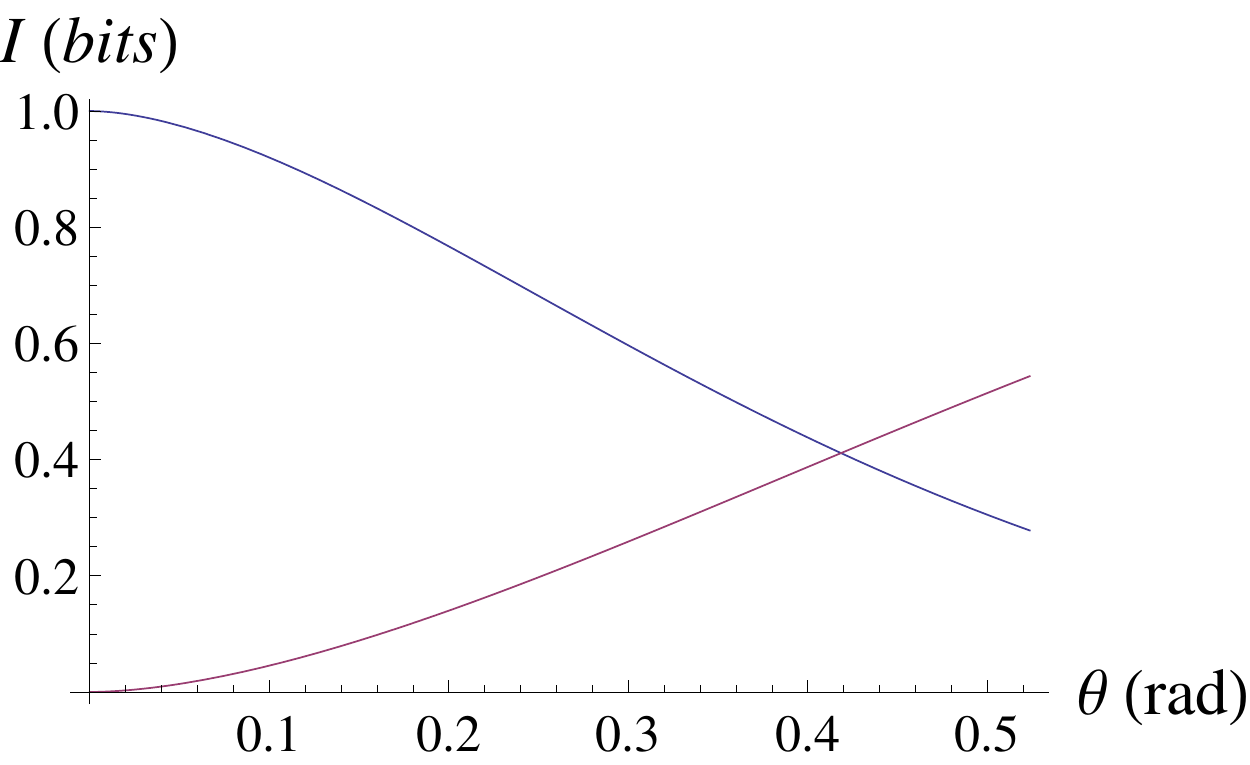}
\caption{The falling  curve represents $I_B = I_C$  (in this symmetric
  model,  where  Eve  attacks   both  arms  with  the  same  strength,
  parametrized by $\theta$), while the rising curve represents $I_E$.}
\label{fig:plot}
\end{center}
\end{figure}

Let us consider the disturbance caused by Eve.  In the given direction
of  polarization   of  the  photon,  Alice's  beam   splitter  may  be
represented as:
\begin{eqnarray}
d_1^\dag &=& \frac{1}{\sqrt{2}}(b^\dag + i c^\dag) \nonumber \\
d_2^\dag &=& \frac{1}{\sqrt{2}}(b^\dag - i c^\dag),
\end{eqnarray}
where $a^\dag,  b^\dag$ are the  creation operators for the  modes $A,
B$, respectively, and $d_1^\dag$ and $d_2^\dag$ are creation operators
corresponding to  detections at $D_1$ and  $D_2$, respectively. Hence,
the state $|\phi\rangle_{AB}$ evolves to
\begin{eqnarray}
|\Psi^\prime\rangle       &\rightarrow&       \frac{1}{\sqrt{2}}\left(
\frac{(|D_1\rangle    +    |D_2\rangle)_{BC}}{\sqrt{2}}|y,n\rangle_{E}
\right.     \nonumber   \\    &    +&\left.    \frac{(|D_1\rangle    -
  |D_2\rangle)_{BC}}{\sqrt{2}}|n,y\rangle_{E}\right),
\label{eq:prob}
\end{eqnarray}
from which, it follows that
\begin{eqnarray}
\textrm{Prob}(D_2|FF)       &=&      \frac{1}{4}|||y,n\rangle_E      -
|n,y\rangle_E||^2 \nonumber \\ &=& \frac{1}{2}\sin^2(\theta).
\label{proba}
\end{eqnarray}
We thus find that the visibility (\ref{eq:visi0}), conditioned on both
applying $F$, falls from 1 to
\begin{equation}
\mathcal{V} = \frac{1+\cos(2\theta)}{2},
\label{eq:visi}
\end{equation}
where,  by  the  assumption  of  channel  losslessness,  $P(D_1|FF)  +
P(D_2|FF)=1$.   The error rate  $e$ in  Eq.  (\ref{eq:e})  becomes, by
Bayesian rule,
\begin{eqnarray}
e = P(FF|D_1)  =  \frac{P(D_1|FF)P(FF)}{P(D_1)} \nonumber \\
= \frac{\sin^2(\theta)}{1+\sin^2(\theta)}  
\label{eq:ee}
\end{eqnarray}
so that the mutual information between Bob and Charlie is
\begin{equation}
I_{BC} = 1 - H(e).
\label{eq:IBC}
\end{equation}
The condition for positive key rate in the protocol is \cite{CK78}
\begin{equation}
K = I_{BC} - \min\{I_{BE}, I_{CE}\} > 0,
\label{eq:+}
\end{equation}
where  $K$ is  the secret  bits that  can be  distilled after  Bob and
Charlie  perform key  reconciliation and  privacy  amplification.  The
security  condition (\ref{eq:+})  becomes,  from Eqs.   (\ref{eq:IE}),
(\ref{eq:ee}) and (\ref{eq:IBC}),
\begin{equation}
H\left(\frac{1-\cos(2\theta)}{4}\right) + 
H\left(\frac{\sin^2(\theta)}{1+\sin^2(\theta)}\right) < 1,
\label{eq:security}
\end{equation}
or $\theta  \lesssim 0.42$ rad,  which, in view of  Eq. (\ref{eq:ee}),
implies $e \lesssim 14.25\%$ (see Fig. \ref{fig:plot}).

\section{Discussion and conclusions \label{sec:discu}}

Here we have extended the concept of counterfactual cryptography
to the  multipartite scenario,  by introducing the  task which  is the
quantum version of CA.  We  have analyzed its security against general
incoherent  attacks.   A   practical  implementation  of  the  present
protocol is  feasible, given the existing  experimental realization of
counterfactual  QKD \cite{RWW+11,  XW11, XST+11,  JSL11,BCD+12}.  CQCA
can  also  be derived  from  the N09  protocol,  just  as the  present
protocol is derived from  the semicounterfactual QKD protocol of Ref.
\cite{SSS13}. The  latter offers a practical advantage  over the former
in that it does not  use polarization encoding, unlike the former.  We
remark that a \textit{non}counterfactual  quantum CA scheme can
be  obtained  using  two-particle  entanglement  and  the  idea  of  a
cryptographic switch  \cite{SOS+12}.  It will be  interesting to study
the security  of such  a protocol, as  compared with the  present CQCA
scheme.

Finally, the above  protocol for CQCA is, as  noted, counterfactual in
the sense  that one  of the two  coplayers transmits  information via
interaction-free measurement, but not both.  Thus, Eve has full access
to Alice's photon, and  the relationship between counterfactuality and
security appears to be less strong than in the Noh protocol.  It would
be  an  interesting  open  problem  to  find  a  multipartite  quantum
cryptographic protocol that is counterfactual in this latter sense.

\bibliography{axta}

\end{document}